\pdfoutput=1      

\documentclass[prl,twocolumn,showpacs,amsmath,amssymb,superscriptaddress]{revtex4}

\usepackage{epsfig}
\usepackage[utf8]{inputenc}
\usepackage{color}

\newcommand{\un}[1]{{\,\text{#1}}}
\newcommand{\vg}{\ensuremath{V_\text{g}}}
\newcommand{\vsd}{\ensuremath{V_\text{SD}}}
\newcommand{\didv}{\ensuremath{\mathrm{d}I/\mathrm{d}\vsd}}
\newcommand{\nh}{\ensuremath{N_\text{h}}}
\newcommand{\kb}{\ensuremath{k_\text{B}}}

\begin{document}

\title{Pumping of vibrational excitations in a Coulomb blockaded
suspended carbon nanotube}

\author{A. K. H\"uttel}
\email{andreas.huettel@physik.uni-regensburg.de}
\altaffiliation[Present address: ]{%
Institute for Experimental and Applied Physics, University of
Regensburg,  93040 Regensburg, Germany.
}%
\affiliation{%
Molecular Electronics and Devices, Kavli Institute of Nanoscience, 
Delft University of Technology, PO Box 5046, 2600 GA Delft, The Netherlands
}%
\author{B. Witkamp}
\affiliation{%
Molecular Electronics and Devices, Kavli Institute of Nanoscience, 
Delft University of Technology, PO Box 5046, 2600 GA Delft, The Netherlands
}%
\author{M. Leijnse}
\author{M. R. Wegewijs}
\affiliation{%
Institut f\"ur Theoretische Physik A, RWTH Aachen, 52056 Aachen, Germany
}
\affiliation{%
Institut f\"ur Festk{\"o}rper-Forschung - Theorie 3,
Forschungszentrum J{\"u}lich, 52425 J{\"u}lich,  Germany
}
\affiliation{%
JARA - Fundamentals of Future Information Technology
}%
\author{H. S. J. van der Zant}
\affiliation{%
Molecular Electronics and Devices, Kavli Institute of Nanoscience, 
Delft University of Technology, PO Box 5046, 2600 GA Delft, The Netherlands
}%

\date{\today}

\begin{abstract}
Low-temperature transport spectroscopy measurements on a suspended
few-hole carbon nanotube quantum dot are presented, showing a
gate-dependent harmonic excitation spectrum which, strikingly, occurs in the
Coulomb blockade regime.
The quantized excitation energy corresponds to the scale expected for
longitudinal vibrations of the nanotube.
The electronic transport processes are identified as
cotunnel-assisted sequential tunneling, resulting from non-equilibrium
occupation of the mechanical mode.
They appear only above a high-bias threshold at the scale of electronic
nanotube excitations.
We discuss models for the pumping process that explain the
enhancement of the non-equilibrium occupation and show that it is connected to a subtle 
interplay between electronic and vibrational degrees of freedom.
\end{abstract}


\maketitle

The coupling of vibrational modes and electronic transport in nanoscale systems
and in particular quantum dots is currently the focus of many theoretical
\cite{prb-braig:205324, prb-luffe:125306} and experimental
\cite{nature-sazonova:284, nl-witkamp:2904, nature-leroy:371,
prl-sapmaz:026801, nature-park:57, nature-smit:906, arxiv-leturcq:08123826}
research efforts. In this respect, single wall carbon nanotubes (SW-CNTs)
provide a unique mesoscopic system where both bulk beam mechanics
\cite{nature-sazonova:284, nl-witkamp:2904} and quantization of phonon modes
\cite{nature-leroy:371,prl-sapmaz:026801,arxiv-leturcq:08123826} have already
been demonstrated. Low-temperature Coulomb blockade (CB) spectroscopy on quantum
dots
formed within the nanotube \cite{science-bockrath:1922, nature-tans:474} has led
to a well-developed understanding of the electronic structure of SW-CNTs
\cite{nphys-deshpande:314, nature-kuemmeth:448, nature-jarillo:389}. In 
suspended SW-CNT quantum dots, transport spectroscopy has also revealed the
Franck-Condon effect \cite{prb-braig:205324}, where the quantized vibrations of
the nanotube become
visible in single electron tunneling (SET) at finite bias. A large electron
phonon coupling \cite{prl-sapmaz:026801, arxiv-leturcq:08123826}, and first
indications of vibrational phenomena in cotunneling have been observed
\cite{arxiv-leturcq:08123826, quantumnems}.

In this Letter we present low-temperature transport measurements on a suspended
CNT quantum dot system. We observe signatures of non-equilibrium population of
the quantized mechanical oscillations in the transport spectrum, revealed by
cotunnel-assisted sequential electron tunneling (CO-SET). The non-equilibrium
occupation is enhanced (``pumped'') by higher-order tunnel processes.
Detailed models are discussed which explain the observations.

The basic device geometry  is sketched in
%
\begin{figure}[th]\begin{center}
\epsfig{file=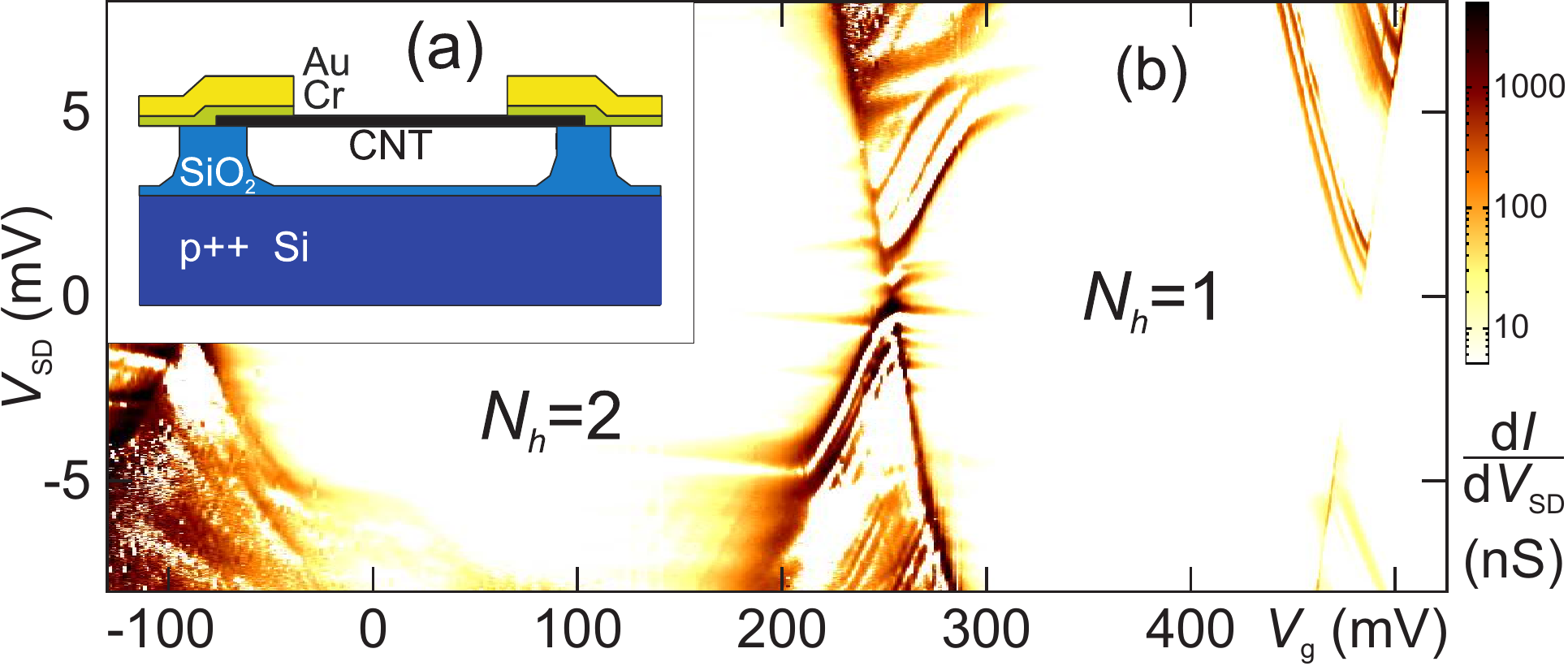, width=8cm}
\end{center}
\vspace*{-0.3cm}
\caption{
(Color online)
(a) Side view drawing of the sample geometry. (b) Overview measurement of the
differential conductance \didv\ of the nanotube quantum dot system in the
few-hole region ($1\le \nh \le 2$, with $\nh=0$ and $\nh=3$ visible at the
edges of the plot) as function of back gate voltage \vg\ and
bias voltage \vsd\ (logarithmic color scale, negative differential
conductance is plotted white). In the CB regions, the
number of trapped valence band holes \nh\ is indicated.
}
\label{fig-overview}
\end{figure}
%
Fig.~\ref{fig-overview}(a). A highly p${}^{++}$ doped silicon wafer, also
functioning as back gate, with $500\un{nm}$ thermally grown oxide on top
provides the starting point of the sample preparation. After lithographic
fabrication of marker structures and localized deposition of growth
catalyst \cite{nature-kong:878}, carbon nanotubes are grown in situ by chemical
vapor deposition and located using atomic force microscopy. Contact
electrodes consisting of $5\un{nm}$ chromium and $50\un{nm}$ gold are deposited
on top of the nanotubes. In a last step, the nanotube devices are suspended
using wet etching in buffered hydrofluoric acid. 

Measurements were performed in a dilution refrigerator with a base temperature
$T_\text{MC}\lesssim 20\un{mK}$ and an electron temperature $T_\text{el} \simeq
100\un{mK}$. Fig.~\ref{fig-overview}(b) shows the differential conductance
\didv\ as function of back gate (substrate) voltage \vg\ and bias voltage \vsd\
of one particular suspended nanotube quantum dot. This sample has a
lithographically designed length of $l=250\un{nm}$, and was -- from transport
measurements as in Fig.~\ref{fig-overview}(b) -- identified to be a 
semiconducting nanotube with a band gap of $E_g=200\un{meV}$. 
Band gap energy and device length lead to a predicted energy scale of $\Delta
E\simeq 6\un{meV}$ of orbital electronic excitations \cite{nature-jarillo:389}.
In the measurement of Fig.~\ref{fig-overview}(b), the basic structure of
diamond-shaped CB regions with a fixed trapped charge, as expected
for a single quantum dot, is clearly visible. Since the band gap position is
known, we can identify the charge states  as $\nh=1$ and $\nh=2$, respectively.

\begin{figure}[th]\begin{center}
\epsfig{file=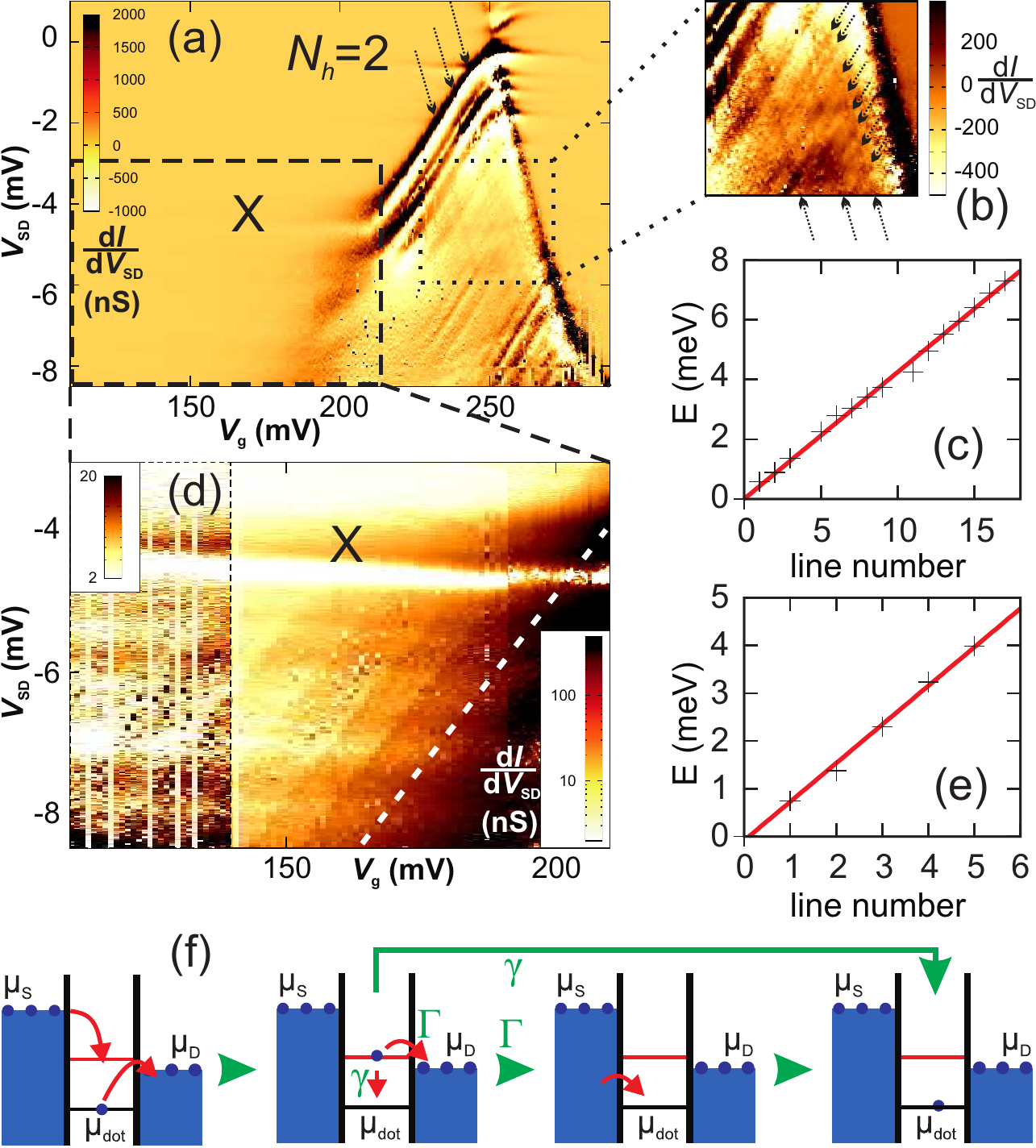, width=8cm}
\end{center}
\vspace*{-0.3cm}
\caption{
(Color online)
(a) Differential conductance \didv\ in the region where the nanotube is charged
with $1\le \nh \le 2$ holes, as function of gate voltage \vg\ and bias voltage
\vsd\ (linear color scale). 
(b) Detail zoom of part of the SET region with $1\le \nh \le 2$, as marked in
(a), using a different color scale. In both (a) and (b), arrows point out line
features corresponding to excited states
(see text) \cite{supp-info}.
(c) Excitation energies corresponding to the lines of enhanced \didv\ in (a) and
(b) with positive slope; the x-axis is the line number.
The solid line is a linear fit, resulting in an average energy difference of
$\Delta E=0.425\pm 0.004\un{meV}$ per line.
(d) Detail zoom of the Coulomb blockade region with $\nh=2$, as marked
in (a) (logarithmic color scale) \cite{supp-info}. A white
dashed line sketches the edge of the SET region. For better contrast,
different color scale ranges are chosen in parts of the plot.
(e) Relative excitation energies for the
line features in (d), using the same parameters for position to energy
conversion. A linear fit results in $\Delta E = 0.810 \pm 0.025\un{meV}$.
(f) Schematic detailing cotunnel-assisted sequential tunneling (CO-SET) (see
text). 
}
\label{fig-12holes}
\end{figure}

Figure~\ref{fig-12holes}(a) 
%
%
displays a detail measurement of the single electron tunneling region with
$1 \le \nh \le 2$ at low
negative bias, and the adjacent $\nh=2$ CB region. A rich spectrum
of equidistant excitation lines with positive slope, corresponding to
excitations of the $\nh=1$ system (see e.g. Ref.~\cite{nature-jarillo:389}), is
found in SET \cite{supp-info}. This is also detailed in the enlarged plot of
Fig.~\ref{fig-12holes}(b) (see arrows).  Figure~\ref{fig-12holes}(c) shows the
corresponding
excitation energies as function of line number \footnote{No clear features can
be identified in the data at the positions expected for lines 4 and 10. This may
e.g. be due to electronic transitions shadowing the vibrational effects in
transport.}. We assign these excitations to a harmonic vibration whose frequency
is $\hbar\omega = 0.425\un{meV} \pm 0.004\un{meV}$, in good agreement with the
bulk mechanics prediction of $\hbar\omega_\text{vib}=0.44\un{meV}$ for the
longitudinal vibration mode of a $250\un{nm}$ long nanotube segment
\cite{prl-sapmaz:026801}. In addition, the data of Fig.~\ref{fig-12holes}(a-b)
reveals three faint excitation lines with negative slope (i.e. $\nh=2$
excitations), marked by three black arrows and
separated by approximately $0.7\un{meV}$. It is difficult to confirm a harmonic
spectrum because of the faintness of the three lines. 

Figure~\ref{fig-12holes}(d) enlarges the region outlined
in Fig.~\ref{fig-12holes}(a) by a black dashed rectangle, plotting the
differential conductance in logarithmic color scale. Here, CB
stabilizes a total charge of $\nh=2$ holes on the nanotube, suppressing SET.
Surprisingly, a pattern of gate dependent excitation lines in the CB
region, parallel to the edge of the SET region, emerges
\cite{supp-info}. Their relative excitation energies are plotted in
Fig.~\ref{fig-12holes}(e). A regular spacing corresponding to a harmonic
oscillator energy of $\hbar \omega = 0.810\un{meV} \pm 0.025\un{meV}$ is
visible, close to the above mentioned energy scale $0.7\un{meV}$ of the $\nh=2$
excitations in SET.

The most straightforward explanation for these lines is that they correspond to
cotunnel-assisted sequential tunneling (CO-SET) processes
\cite{prb-schleser:206805}. These are multi-step processes, as sketched in
Fig.~\ref{fig-12holes}(f). Inelastic cotunneling leaves a quantum dot in an
excited state (see second panel in the figure), which is possible at any voltage
above the energy of this state. Subsequently, sequential tunneling to the drain
electrode can follow for voltages close to the Coulomb diamond edges, even if
the Coulomb-blockaded ground state does not allow it. For this to happen, the
rate $\Gamma$ for tunneling out has to be comparable to or larger than the
energy relaxation rate $\gamma$ to the ground state.  

In general, CO-SET processes can involve intrinsic (e.g. orbital) excitations of
quantum dots as well as vibrational states. Here, as opposed to the measurements
of Ref.~\cite{prb-schleser:206805}, the highly regular excitation spectrum
observed in Fig.~\ref{fig-12holes}(d) strongly indicates a vibrational origin.
The vibrational CO-SET lines are related to the two-hole charge state
vibrational excitations seen in SET with negative slope. As the vibrational
energy exceeds thermal broadening ($\hbar \omega \gg \kb T$), the observation of
multiple excitations in CB strongly suggests storage and 
subsequent release of energy in the vibrational mode, involving phonon
absorption sidebands \cite{prb-luffe:125306, Leijnse08theory, nature-leroy:371}.
This stands in contrast to the usual assumption that vibrational relaxation is
fast and that the mechanical system is predominantly found in 
its ground state. The required energy is provided by inelastic cotunneling,
exciting the vibrational mode and causing its non-equilibrium occupation.

A particularly interesting feature of the measurement of
Fig.~\ref{fig-12holes}(a-e) is that the CO-SET lines can only be observed beyond
a
weakly gate-dependent threshold at $\vsd \simeq -4\un{mV}$ (marked X in the
plots), where the differential conductance is enhanced; i.e. they only occur at
energies higher than $4\un{meV}$.
%
\begin{figure}[th]\begin{center}
\epsfig{file=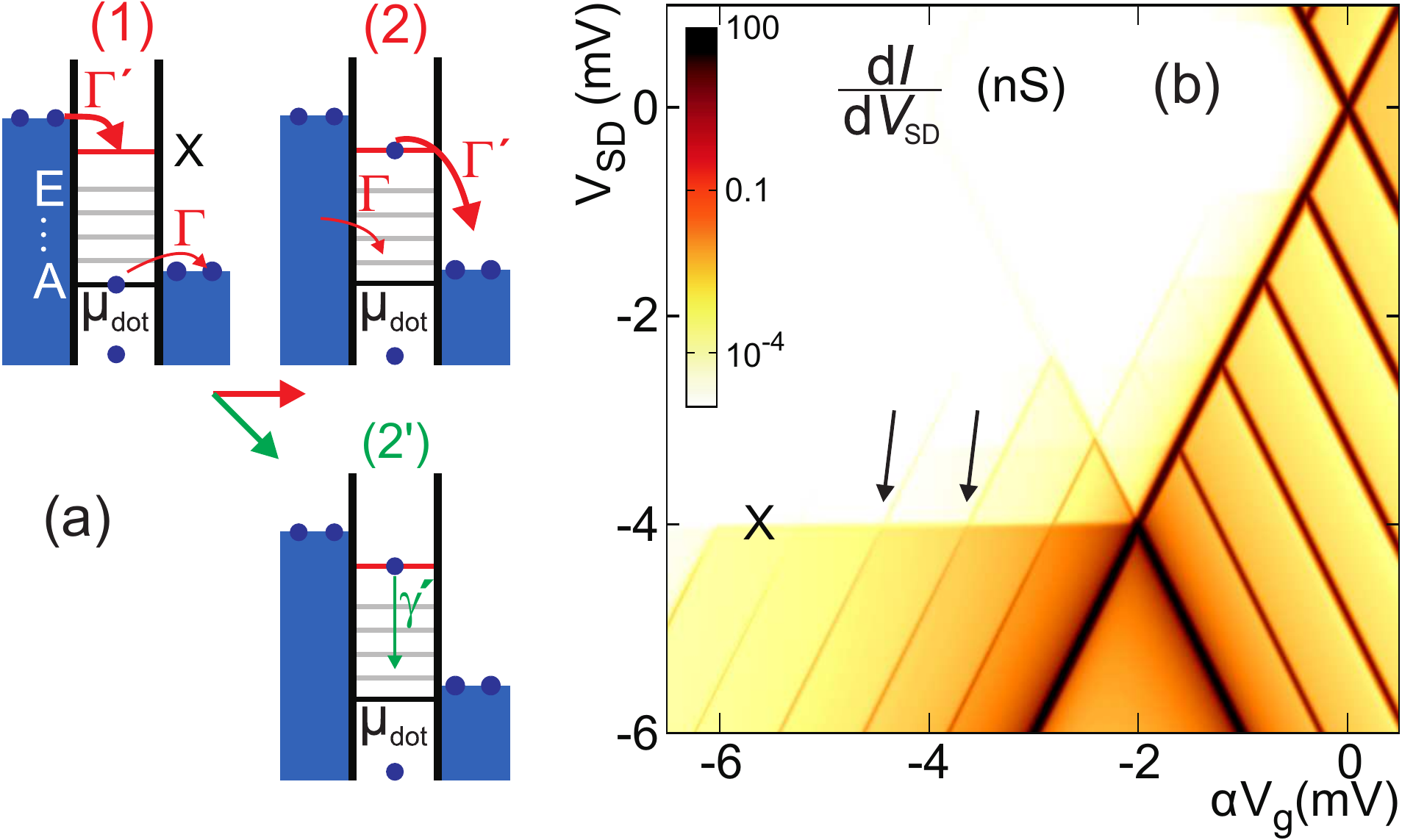, width=8cm}
\end{center}
\vspace*{-0.3cm}
\caption{
(Color online) Single quantum dot model reproducing the observed transport
features: 
(a) Process and energy level scheme analogous to Fig.~\ref{fig-12holes}(f),
leftmost panel, showing the lead Fermi levels, the accessible states
within the quantum dot, and tunneling and relaxation rates relevant for
pumping the vibration mode into non-equilibrium \cite{supp-info}. For a detailed
description of the panels and the rates $\Gamma$, $\Gamma'$,
and $\gamma'$ see the text.
(b) Calculated differential conductance as function of gate voltage \vg\ and
bias voltage \vsd, for the following parameters: vibrational level-spacing
$\hbar \omega=810 \mu \text{eV}$, $kT = 8.6 \mu \text{eV} = 10 \hbar \Gamma' =
10^3 \hbar \Gamma = 10^4 \hbar \gamma' = 10^4 \hbar \gamma$. The tunnel
couplings were chosen smaller than in the experiment to ensure a well-behaved
perturbation expansion. \vg\ is scaled with the gate conversion factor
$\alpha=C_\text{g}/C$ (see e.g. \cite{quantumnems,prb-schleser:206805}). Arrows 
indicate the enhancement of CO-SET at high bias.
}
\label{fig-process}
\end{figure}
%
The line feature marked with X in the measurements represents the onset of an
inelastic cotunneling current, 
corresponding to an electronic excitation of the quantum dot. To understand the
resulting interplay between mechanical and electronic excitations we have
performed model calculations, illustrated in Fig.~\ref{fig-process}. The
transport calculations account for strong Coulomb effects as well as tunneling
processes up to 4th order in the tunneling Hamiltonian, responsible for
cotunneling, line broadening and level renormalization which are important close
to SET resonance; see Ref.~\cite{Leijnse08theory} for details.

The two-hole ground state and its vibrational excitations are the origin of
CO-SET lines as observed in the experiment. To demonstrate this idea, it
suffices to consider a model with several equidistant states A--E (see 
Fig.~\ref{fig-process}(a), panel (1), and also \cite{supp-info}), 
which are coupled with the same rate $\Gamma$ to the one-hole ground state.
In addition, an excited two-hole state X supporting a current with rate
$\Gamma'$ is introduced. Since the edge of the CB diamond becomes very
broad in the experiment beyond this excitation, an estimate of the ratio of
$\Gamma$ and $\Gamma'$ based on the SET current is difficult. However, the
strong signature of the inelastic cotunneling line associated with state X 
indicates that $\Gamma' \gg \Gamma$. The result of a calculation based on this
model is shown in Fig.~\ref{fig-process}(b). The important characteristic of the
experiment is reproduced: CO-SET lines are strongly enhanced beyond the
threshold for
inelastic cotunneling connected to state X (see arrows in
Fig.~\ref{fig-process}(b)).

The appearance of a set of harmonic CO-SET lines at high bias can be explained
by a subtle interplay of electronic and vibrational excitations, with two
coexisting paths. Figure \ref{fig-process}(a) illustrates these two paths; 
panels (1) and (2) for one and panel (1) and (2’) for the other. For energies
larger than the one of the inelastic cotunneling step at $\vsd \simeq -
4\un{mV}$ a process involving two stages of inelastic cotunneling becomes
feasible: the first effectively excites the dot from state A to state X (panel
(1)), the second results in a transition from state X into one of the states
B-E (panel (2)). The rate for each such cotunnel process is $\propto \Gamma
\Gamma'$. Compared to the low bias situation, this significantly enhances the
population of the vibrational excited states B-E, and thereby the visibility of
the vibrational CO-SET lines, since without access to state X, the limiting
rates for CO-SET are proportional to $\Gamma^2 << \Gamma \Gamma'$. Thus, while
CO-SET lines can in principle also be present at small bias ($\left| \vsd \right| < 4\un{mV}$),
beyond the inelastic contunneling step they appear more pronounced: The strongly
coupled excited state pumps the vibrational mode out of equilibrium, enhancing
its population. 

The second path that increases the occupation of the vibrationally excited 
states B-E involves direct relaxation of state X (panel (2') in 
Fig.~\ref{fig-process}(a)) with a rate $\gamma'$. As the precise value of
$\gamma'$ is unknown, we have verified that the qualitative result that the
CO-SET lines are strongly enhanced beyond the inelastic cotunneling step
connected to state X persists for a large range of this parameter. 
An important further qualitative conclusion from the observation of CO-SET
features is that relaxation from states B-E into the two-hole ground state A
(rate $\gamma$ in Fig. \ref{fig-12holes}(f)) in the experiment can not be faster
than tunneling out of the quantum dot ($\gamma \lesssim \Gamma$). 

%
\begin{figure}[th]\begin{center}
\epsfig{file=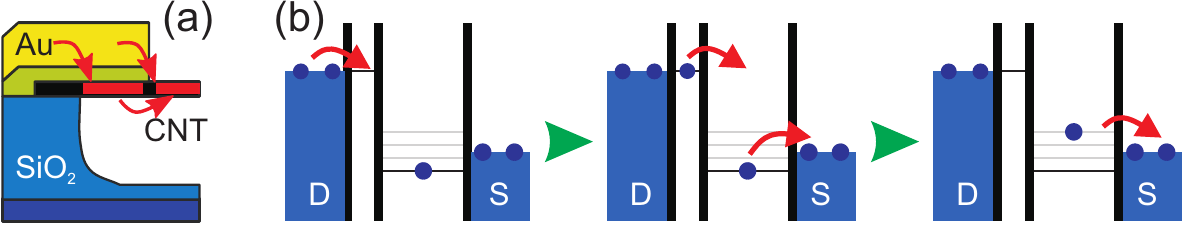, width=8cm}
\end{center}
\vspace*{-0.3cm}
\caption{
(Color online) Alternative mechanism for obtaining a vibrational non-equilibrium
occupation.
(a) Schematic side-view drawing detailing the possible formation of a small
quantum dot beneath one of the metallic contact leads. 
(b) Transport processes for this `double quantum dot' case: additional 
cotunneling processes are enabled when the small quantum dot enters the energy
window given by \vsd\ (see text).
}
\label{fig-double}
\end{figure}
%
So far we have assumed that the weakly gate-dependent feature X in 
Fig.~\ref{fig-12holes} corresponds to an electronic {\it excited} state. An
alternative explanation invokes an electronic {\it ground} state of a second
quantum dot forming within the nanotube, e.g. beneath one of the contact
electrodes, as drawn in Fig.~\ref{fig-double}(a). Assuming this state has a
much weaker gate coupling, we arrive at an equivalent mechanism for the
non-equilibrium occupation of the vibrational mode. Local distortions and
curving edges in the measurement of Figs.~\ref{fig-overview} and
\ref{fig-12holes} may be taken as a hint of such a scenario, as has also been
reported in Ref.~\cite{nl-grove-rasmussen:1055}.

Fig.~\ref{fig-double}(b) illustrates the transport process for this `double
quantum dot' case. The shielded additional small quantum dot displays a much
weaker gate voltage dependence of the level energy. The line of enhanced
differential conductance marked in Fig.~\ref{fig-12holes} with X now describes
its alignment at the drain lead with the drain Fermi edge. The small quantum dot
can then be occupied by a sequential tunneling process as sketched in
Fig.~\ref{fig-double}(b). Inelastic cotunneling through the main dot may
follow, leading to the occupation of an excited vibrational state. Although the
origin of, and tunneling processes associated with, state X differ from the
first scenario, the main effect is the same: state X pumps the vibrational mode
through consecutive tunneling events. 

An intriguing feature of the measurement that has not been discussed so far is
that the vibrational frequency in the $\nh=2$ charge state, as measured both
from SET and CO-SET lines, is approximately twice as large as that in the
$\nh=1$ charge state, seen as SET features only. This difference does not affect
the pumping mechanism. It is possibly related to the fact that the observed
quantum dot is in the few-carrier regime -- in contrast to previous work on
longitudinal phonon excitations in carbon nanotubes \cite{prl-sapmaz:026801}.
Whereas tension only affects the longitudinal mode via higher-order effects
\cite{prl-sapmaz:026801}, the transition from e.g. one to two trapped holes
involves a distinct spatial redistribution of charge along the nanotube
\cite{nphys-deshpande:314}. This strongly affects the electrostatic force
distribution, which makes variations of vibration mode energies and mode shapes
likely. As shown in Ref.~\cite{prl-sapmaz:026801}, the coupling to the lowest
symmetric vibrational mode of the nanotube may vanish for a charge distribution
localized at the center of the nanotube. In general one may thus expect 
that the excitation of different modes depends strongly on the charge state, as
we observe in the experiment. 

Finally, the experimental observation of vibrational CO-SET resonances allows to
establish a lower boundary for the quality factor of the longitudinal mechanical
mode \cite{prb-luffe:125306}, independent of the detailed 
pumping mechanism. The tunnel current at the edge of the low-bias CB region
($\left|\vsd\right| < 4\un{mV}$), $I\simeq 1\un{nA}$, provides an approximation
for the SET tunneling rate $\Gamma$. For CO-SET to
be visible, the lifetime of vibrational excitations must be larger than the
corresponding timescale $\tau \simeq 0.16\un{ns}$. With the vibrational energy
$\hbar \omega=810\,\mu\text{eV}$, one obtains $Q \gtrsim \tau\times \omega/2\pi
= 31$. Since we observe several sidebands, a higher value for $Q$ is more likely
than this lower boundary, depending on the specific relaxation processes. 

In conclusion, we have observed a pattern of equidistant, gate-dependent
vibrational excitations \emph{in the Coulomb blockade regime} of a suspended
carbon nanotube quantum dot. The appeareance of the excitations is explained in
the context of co-tunnel assisted sequential electron tunneling via phonon
absorption processes. Interestingly, the absorption sidebands of the quantized
phonon mode are only visible above a finite bias voltage threshold. Two models
for the enhancement of the non-equilibrium distribution of the mechanical mode
are discussed which demonstrate that the mode can be ``pumped'' either by an
electronic excitation of the nanotube, or by a ground-state of another small
quantum dot within the same nanotube. The pumping mechanism offers a perspective
for electric control of quantized mechanical motion in nanoscale transistors by
employing the interplay with electronic degrees of freedom.

The authors would like to thank Y. Blanter and M. Poot for
insightful discussions, and B. Otte and H. Pathangi for experimental help.
Financial support by the Dutch organization for Fundamental Research on Matter
(FOM), the NWO VICI program, NanoNed, DFG SPP-1243, the Helmholtz Foundation,
and the FZ-J\"ulich (IFMIT) is acknowledged.

\bibliography{paper}

\end{document}